\documentclass{PoS}

\title{Group-theoretical construction of finite-momentum and multi-particle
operators for lattice hadron spectroscopy}

\ShortTitle{Finite-momentum and multi-particle operators for hadron spectroscopy}

\author{\speaker{Justin Foley},$^a$ John Bulava$^b$, You-Cyuan Jhang$^c$, Keisuke J. Juge$^d$, 
  David Lenkner$^c$, Colin Morningstar$^c$ and Chik Him Wong$^e$\\
        \llap{$^a$} Department of Physics and Astronomy, University of Utah, Salt Lake City, UT 84112, USA\\
        \llap{$^b$}CERN, Physics Dept. 1211 Geneva 23, Switzerland \\
        \llap{$^c$}Dept. of Physics, Carnegie Mellon University, Pittsburgh, PA 15213, USA \\
        \llap{$^d$}Dept. of Physics, University of the Pacific, Stockton, CA 95211, USA \\
        \llap{$^e$}University of California San Diego, La Jolla, CA 92093, USA \\
	Email: \email{jfoley@physics.utah.edu}}

\abstract{
\vspace*{-10.5cm}
\begin{flushright}
\texttt{\footnotesize CERN-PH-TH/2012-130}\\
\end{flushright}
\vspace*{9.5cm} 
Determining the spectrum of hadronic excitations from Monte Carlo simulations requires the use of interpolating operators that couple to multi-particle states. Recent algorithmic advances have made the inclusion of multi-hadron operators in spectroscopy calculations a practical reality. In this talk, a procedure for constructing a set of multi-hadron interpolators that project onto the states of interest is described. To aid in the interpretation of simulation data, operators are designed to transform irreducibly under the lattice symmetry group. The identification of a set of optimal single-hadron interpolators for states with non-zero momenta is an essential intermediate step in this analysis.}

\FullConference{ The XXIX International Symposium on Lattice Field Theory - Lattice 2011\\
July 10-16, 2011\\
Squaw Valley, Lake Tahoe, California}

\begin{document}

\subsection*{Introduction}
The elucidation of the spectrum of hadronic excitations
predicted by QCD is a long-term goal of lattice field theory. 
While a number of low-lying stable-particle masses are 
readily accessible, most of the low-lying spectrum has, until recently, 
lain beyond reach. The evaluation of correlation functions containing
disconnected quark-line diagrams, which arise in the isosinglet-meson 
sector, is extremely challenging. Multi-particle correlators, needed to 
study states above threshold, are, in general, even more problematic.

The stochastic LapH method, described in Ref.~\cite{laph}, is an efficient algorithm 
for evaluating hadronic correlators involving same-timeslice quark propagation.
This algorithm facilitates the accurate evaluation of hadronic correlation functions using far fewer quark-matrix inversions than earlier methods.
Moreover, using this approach, hadronic two-point functions factorize into 
source and sink contributions, greatly simplifying the evaluation of multi-hadron correlators in particular.

The next step towards understanding the experimentally-accessible spectrum
involves the calculation of the lowest-lying stationary-state energies 
on multiple lattice volumes. In a given symmetry channel, multiple energy levels can be determined from the solutions of a generalized eigenvalue problem involving matrices of hadronic two-point functions~\cite{luscher}. However, the success of this approach depends on the identification of appropriate interpolating operators.
Considerable effort has already been invested in the construction of single-particle interpolators~\cite{baryon,meson}. In this note, we report on progress in constructing the multi-hadron interpolators needed above threshold.

Although we are primarily interested in the spectrum of hadrons at rest, 
two or more hadrons in flight can combine to form a state with zero net momentum. 
Hence, as an intermediate step in our calculations, it is also necessary to construct 
interpolating operators for states with non-zero momenta.
The analysis is further complicated by the fact that different flavor sectors become 
intertwined once multi-particle states are taken into account. For example, an isovector two-meson 
state might have one isosinglet constituent meson.
To make sense of lattice data, it is essential to use interpolating operators 
with well-defined lattice quantum numbers. 
In particular, we use operators that transform irreducibly under the spatial 
symmetry group of the lattice. 
Our simulations are performed with $2+1$ flavors of anisotropic clover fermions, and we are thus able to construct light-hadron operators with definite isospin or G-parity. 
Due to the finite temporal extent of the lattice, 
it is also advantageous to use meson operators that are irreducible under time reversal. Here, we focus on the spatial symmetries only. Incorporating the flavor and time-reversal symmetries is straightforward, and will be described in full in a forthcoming journal publication.

\subsection*{Group theory}
In simulations on a cubic spatial lattice, 
the energies of zero-momentum states are commonly classified according 
to the irreducible representations (irreps) of the octahedral group, denoted $O_{h}$ in 
Sch\"{o}nflies notation.
The zero-momentum irreducible representations 
are a subset of the irreps
of the space group of the cubic lattice, which is 
the semi-direct product of 
$O_{h}$  
and the group of lattice translations $\mathcal{T}$. 
The translation group is Abelian. 
Hence, all of its irreducible representations are one-dimensional. 
They are characterized by a lattice momentum $\mathbf{p}$.
Moreover, $\mathcal{T}$ is invariant under the full space group.
Therefore, the method of induced representations is applicable,
and the full set of space-group irreps can be deduced from the 
irreducible representations of the little groups of the allowed 
lattice momenta.
The little group of the lattice momentum $\mathbf{p}$ is the subgroup 
of lattice rotations which leaves $\mathbf{p}$ invariant.
For mesons at rest, the relevant little group is the octahedral group.
In the continuum, the little group of non-zero momenta is $C_{\infty v}$. 
$C_{n v}$ being the symmetry group of an n-sided regular pyramid.
On the lattice, the little group depends on the orientation 
of the momentum with respect to the lattice axes.
Hence, the little group of momenta along a lattice axis, (0,0,1), for example, 
is $C_{4v}$; for planar-diagonal momenta, in the direction (0,1,1), say, the 
little group is $C_{2 v}$; and for momenta along a cubic diagonal, the 
little group is $C_{3 v}$. 

Periodic spatial boundary conditions are enforced in our simulations, and
irreps for all lattice states with momenta satisfying $ |{\mathbf{p}}| \leq 4 \pi/L$ can be induced from representations 
of the little groups listed above. 
The relevant little group for higher-momentum states moving in the $(0,1,2)$, $(1,1,2)$ and equivalent directions
is $C_{2}$, the two-fold cyclic group. 
For the remainder of this report, we focus on states with momenta 
satisfying $|{\mathbf{p}}| \leq 4 \pi/L$. While we expect multi-particle states involving pions with 
higher
momenta to play a role in the low-lying spectrum in current lattice simulations, the symmetry properties of these states are generally trivial to deduce.

$C_{4 v}$ has five single-valued irreducible representations, 
which we label $A_{1}$, $A_{2}$, $B_{1}$, $B_{2}$ and $E$.
We follow the Mulliken naming convention, so $A$ and $B$  
denote one-dimensional representations, and $E$ is two-dimensional.
$C_{2 v}$  has single-valued irreps $A_{1}$, $A_{2}$, $B_{1}$, $B_{2}$,
and the irreps of $C_{3 v}$ are $A_{1}$, $A_{2}$ and $E$.
The double-valued little-group irreps are one- or two-dimensional.
To be consistent with the usual labeling convention for 
zero-momentum double-valued irreps~\cite{johnson}, we use $G$ to denote 
two-dimensional representations, and $F$ to denote one-dimensional 
fermionic
irreps~\footnote{Our labeling convention differs from the 
convention in Ref.~\cite{fleming}, for example}. 

In the continuum and infinite-volume limits, non-zero momentum irreps are labeled by the 
absolute value of the helicity. The continuum 
representations are two-dimensional 
except for two zero-helicity irreps.
The relationship between the various lattice little-group irreps 
and their continuum counterparts and the irreps of $O_{h}$
are determined by subduction. Table~\ref{table:subduction} contains the 
subduction of the irreps of $O_{h}$ to $C_{4 v}$, $C_{2 v}$ and $C_{3v}$.
The table tells us, for example, that a pion with momentum 
along a lattice axis transforms according to the $A_{2}$ little-group irrep, 
and a vector meson, in the $T_{1 u}$ irrep at rest, can appear in the 
$A_{1}$ (scalar) or $E$ irrep, depending on whether the momentum 
is parallel or perpendicular to the direction of polarization.
\begin{table}
\begin{center}
\begin{tabular}{|c|c|c|c|}
\hline 
$\Lambda \left( O_{h} \right) $ 
&  $\downarrow C_{4 v} $ 
&  $\downarrow C_{2 v} $ 
&  $\downarrow C_{3 v} $ \\
\hline
$A_{1 g}$ & $A_{1}$ & $A_{1}$ & $A_{1}$ \\
$A_{1 u}$ & $A_{2}$ & $A_{2}$ & $A_{2}$ \\
$A_{2 g}$ & $B_{1}$ & $B_{2}$ & $A_{2}$ \\
$A_{2 u}$ & $B_{2}$ & $B_{1}$ & $A_{1}$ \\
$E_{g}$   & $A_{1} \oplus B_{1} $ & $A_{1} \oplus B_{2}$ & $E$ \\
$E_{u}$   & $A_{2} \oplus B_{2} $ & $A_{2} \oplus B_{1}$ & $E$ \\
$T_{1 g}$ & $A_{2} \oplus E $   &  $A_{2} \oplus B_{1} \oplus B_{2}$ & $A_{2} \oplus E$\\
$T_{1 u}$ & $A_{1} \oplus E $   &  $A_{1} \oplus B_{1} \oplus B_{2}$ & $A_{1} \oplus E$\\
$T_{2 g}$ & $B_{2} \oplus E $   &  $A_{1} \oplus A_{2} \oplus B_{1}$ & $A_{1} \oplus E$\\
$T_{2 u}$ & $B_{1} \oplus E $   &  $A_{1} \oplus A_{2} \oplus B_{2}$ & $A_{2} \oplus E$\\
$G_{1 g/u}$ & $G_{1}$  & $G$ & $G$\\ 
$G_{2 g/u}$ & $G_{2}$  & $G$ & $G$ \\
$H_{g/u}$ & $G_{1} \oplus G_{2} $ & $2G$ & $F_{1} \oplus F_{2} \oplus G$ \\
\hline
\end{tabular}
\caption{Subduction of the irreps of $O_{h}$ onto the non-zero momentum little groups
$C_{4 v}$, $C_{2 v}$, and $C_{3 v}$.}
\label{table:subduction}
\end{center}
\end{table}

\subsection*{Single-hadron operator construction}
Each space-group irrep is characterized by a set of lattice 
momenta closed under the action of $O_{h}$, and an irreducible 
representation of the corresponding little group. 
In practice, an operator basis for a space-group irrep is found 
by choosing a reference momentum ${\mathbf{p}}$,
and constructing a set of basis operators for the little group of
${\mathbf{p}}$. 
The complete space-group basis is then obtained by 
applying a set of rotations which generate $\mathbf{p}^{*}$, the star of $\mathbf{p}$, to this initial operator set.

To construct little-irrep bases, we first identify 
gauge-invariant elemental operators with definite momenta and 
flavor quantum numbers. The elementals incorporate different 
gauge-covariant displacements, but their transformation properties 
under lattice rotations are straightforward to deduce.
For example, the general expression for a meson elemental annihilation operator
is
\begin{eqnarray}
\phi^{A B}_{a b} \left( t, \mathbf{p} \right) 
= \sum_{\mathbf{x}} e^{-i \mathbf{p} \mathbf{x} }
\left( 
\tilde{\bar{\psi}}^{(A)}_{a} \tilde{D}^{(n) \dagger}_{i}
\right)
\left( \mathbf{x}, t \right)
\left(
\tilde{D}_{j}^{(n)}
\tilde{D}_{k}^{(n)}
\tilde{\psi}_{b}^{B}
\right)
\left( \mathbf{x}, t \right),
\end{eqnarray}
where $a \left( b \right) $ and 
$A(B)$ are spin and flavor indices, and  
$\tilde{\psi}$ denotes a smeared quark field.
The action of the gauge-covariant displacement operators
is given by
\begin{eqnarray}
\tilde{D}^{(n)}_{k} \phi \left( \mathbf{x}, t \right)
= \tilde{U}_{k} \left( \mathbf{x}, t \right) \cdots 
\tilde{U}_{k} \left( \mathbf{x} + (n-1) \hat{k}, t \right) 
\phi \left( \mathbf{x} + n \hat{k} , t \right),
\end{eqnarray}
where $\tilde{U}$ is a (stout-)smeared link variable.

Under the action of the element $R$ of the little group of 
$\mathbf{p}$, ${\phi}_{\alpha} \left(t, \mathbf{p} \right)$
goes to 
\begin{eqnarray}
U_{R} {\phi}_{\alpha} \left( t , \mathbf{p} \right) U_{R}^{\dagger}
 = \sum_{\beta=1}^{M} {\phi}_{\beta}  \left( t, \mathbf{p} \right)
W_{\beta \alpha} \left(R \right)^{*}.
\end{eqnarray} 
The subset of $N$ elementals $\left \{ {\phi} \right \}$ therefore 
generates a representation of the lattice little group, with matrices $W$, which is in general 
reducible.
Basis operators for the constituent irreps are formed from 
linear combinations of these elementals as follows.
First, given explicit representation matrices for the irrep 
$\Lambda$, $\Gamma^{(\Lambda)}$, we define the $N \times N$ projection 
matrix
\begin{eqnarray}
  P^{\Lambda \lambda}_{\alpha \beta} =  \frac{d_{\Lambda}} {n_{G_{\mathbf{p}}^{D}}}
  \sum_{R \in G_{\mathbf{p}}^{D}} \Gamma^{(\Lambda)}_{\lambda \lambda} \left( R \right)
W_{\beta \alpha} \left( R \right)^{*},
\label{eqn:irrep_project}
\end{eqnarray}
where $\lambda$ labels a row of the $\Lambda$ irrep 
and $d_{\Lambda}$ is the dimension of the irrep.
$n_{G_{\mathbf{p}}^{D}}$ is the order of the group $G_{\mathbf{p}}^{D}$, the double cover of the 
little group of $\mathbf{p}$.
If the irrep $\Lambda$ appears in the elemental representation 
$W$, and the elements of row $\alpha$ of $P^{\Lambda \lambda}$ are not all zero, the linear superposition of elementals 
\begin{eqnarray}
{O}^{\Lambda \lambda}_{P_{\alpha}} \left( t, \mathbf{p} \right)
= 
\sum_{\beta = 1 }^{N} P_{\alpha \beta}^{\Lambda \lambda} 
{\phi}_{\beta} \left( t, \mathbf{p} \right)
\end{eqnarray}
transforms according to the row $\lambda$ of the irrep $\Lambda$. 
The rank of $P^{\Lambda \lambda}$ is equal to the
multiplicity of $\Lambda$ in $W$.
Starting with the operator for row $\lambda$, the other operators 
in the little-irrep basis are given by
\begin{eqnarray}
{O}_{P_{\alpha}}^{\Lambda \mu} \left( t, \mathbf{p} \right)
= 
\frac{d_{\Lambda}} {n_{G_{\mathbf{p}}^{D}}}
\sum_{R \in G_{\mathbf{p}}^{D}} \Gamma^{(\Lambda)}_{\mu \lambda} \left(R \right)^{*}
U_{R} {O}^{\Lambda \lambda}_{P_{\alpha}} \left( t, \mathbf{p} \right)
U^{\dagger}_{R}.
\end{eqnarray}

When constructing space-group operators, neither the reference momentum,
$\mathbf{p}$, nor the rotations used to generate ${\mathbf{p}}^{*}$ are specified uniquely.
However, a different choice of reference momentum or star rotations 
simply amounts to a change of basis, and is of no physical consequence.

\subsection*{Single-hadron operator selection} 
\begin{figure}[t]
\centering
\includegraphics[width=0.9\textwidth]{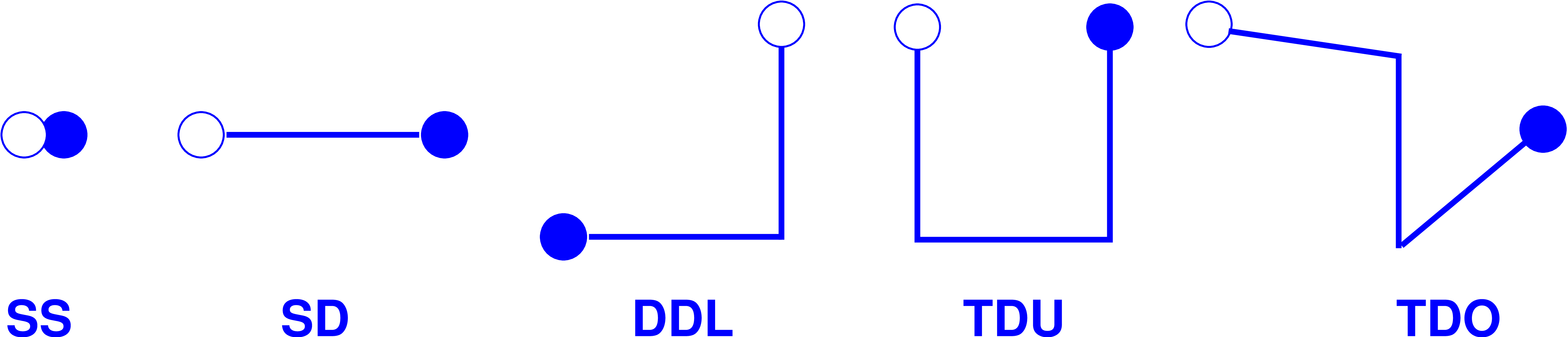}
\caption{Representation of the spatial paths used in the meson 
elemental operators. All five displacement types are used in the 
construction of zero-momentum basis operators. 
Only single-site (SS) and singly-displaced (SD) operators are 
used in the non-zero momentum channels.}
\label{fig:operator}
\end{figure}

In order to project onto different hadronic excitations, 
a number of spatial-displacement combinations are used in the operator 
sets. Fig.~\ref{fig:operator} shows the set of meson-elemental displacements currently in use.
Of these, only the single-site and singly-displaced paths have been used 
in the non-zero momentum single-hadron operators to date. However, 
we are primarily interested in the lightest finite-momentum states, so 
we expect the simplest of the displaced interpolators to suffice.

The operator-construction algorithm described above is readily
automated, and one can easily generate a large number (in some of the zero-momentum irreps, hundreds or even thousands) of irreducible operators.
It is therefore essential to apply a pruning procedure to the 
candidate operator sets to identify 
manageable subsets of clean operators that couple strongly to the 
lowest-lying states in each symmetry channel.
The pruning consists of low-statistics analyses of the spectrum in 
each of the symmetry channels. First, the noisiest of the interpolators 
are identified and discarded. Cross-correlations between the remaining operators are examined in order to identify a subset of interpolators which couples 
well to a number of different states. 

\begin{figure}[t]
\centering
\includegraphics[width=0.7\textwidth]{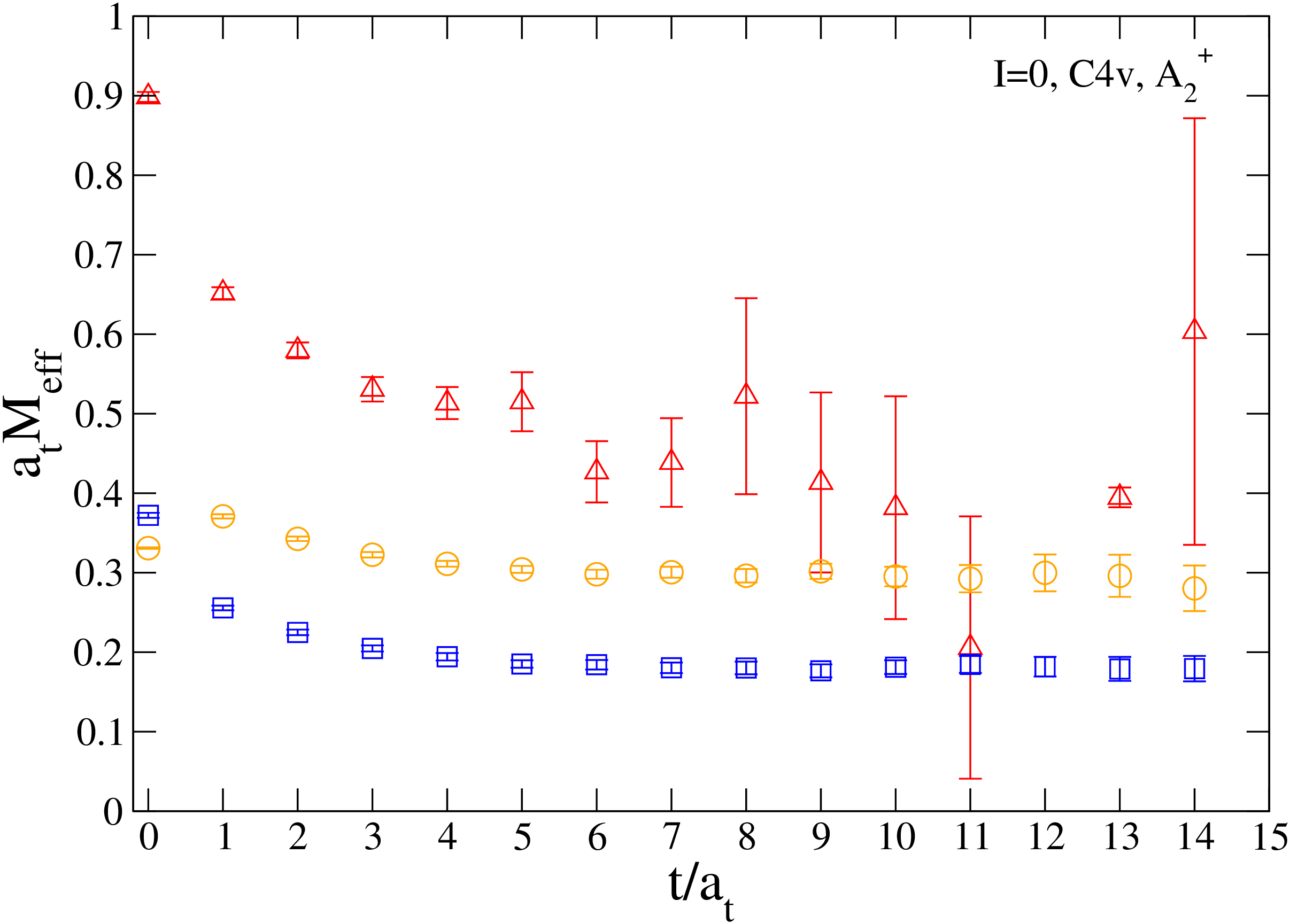}
\caption{Lowest-lying effective masses in an isoscalar meson irrep containing 
states with the minimum on-axis lattice momenta. Effective masses were computed using pruned 
single-meson operators on approximately 100 configurations on a  $(1.9 {\rm{fm}})^{3}$ spatial lattice. The pion mass on these configurations is approximately 400~MeV.
Only single-site and singly-displaced operators were used.
This particular lattice irrep is relevant for the $\eta$ meson.}
\label{fig:moving}
\end{figure}

Effective masses obtained from pruned operators in a non-zero momentum isoscalar meson 
channel can be seen in Fig.~\ref{fig:moving}. These measurements were performed on approximately 100 configurations on a 
$(1.9{\rm{fm}})^{3}$ spatial lattice. These results correspond to the minimum allowed on-axis momenta.

\subsection*{Multi-hadron states}
To form multi-hadron operators, representation matrices for 
the space-group operator bases are needed.
Matrices for non-zero momentum 
irreps are obtained as follows. 
Having chosen a reference momentum $\mathbf{p}$, 
and explicit bases for the little-group irreps $\left \{\Lambda \right \}$ of $\mathbf{p}$, with matrices 
$\Gamma^{\left(\Lambda \right)}$, say,
one identifies a subset of lattice rotations to generate $\mathbf{p}^{*}$, which we 
label $R^{*}$. The components of the induced representation matrices
are
\[
\Gamma^{\left(\Lambda; \mathbf{p}^{*} \right)}_{\left(\alpha, \mathbf{p}_{1}\right), \left( \beta, \mathbf{p}_{2} \right)} \left( R \right)
= \left\{
\begin{array}{ll}
\Gamma^{ \left( \Lambda \right) }_{\alpha \beta} \left( R_{2}^{-1} R R_{1} \right); &~~~\mathbf{p}_{2} = R \mathbf{p}_{1}
\\
0; &~~~\mathbf{p}_{2} \neq R \mathbf{p}_{1},
\end{array}
\right.
\]
where $\mathbf{p}_{1}$ and $\mathbf{p}_{2}$ are momenta in $\mathbf{p}^{*}$, and $R_{i}$ is the rotation in $R^{*}$ that 
takes the reference momentum $\mathbf{p}$ to $\mathbf{p}_{i}$. Hence if $\mathbf{p}_{2} = R \mathbf{p}_{1}$, $R_{2}^{-1} R R_{1}$ 
is a rotation in the little group of $\mathbf{p}$.


The procedure for constructing irreducible multi-hadron 
operators by combining irreducible single-particle operators
is similar to the method used to construct  
single-hadron operators described previously.
The direct products of single-particle irreps form 
(generally reducible) representations of the space group.
These representations can be written in 
block-diagonal form, each block corresponding to a different 
relative orientation of the constituent single-particle momenta.
To make two-particle operators for the rest spectrum,
we need only consider the block with back-to-back single-hadron momenta, which 
itself forms a representation of $O_{h}$.
Here, we use the notation
$\left[ \Lambda_{1}, \Lambda_{2}; \mathbf{p}^{*} \right]$
to denote this representation, where $\Lambda_{1}$ and $\Lambda_{2}$ 
are irreps of the little group of $\mathbf{p}^{*}$, the star of the single-particle momenta. 
The dimension of this two-particle representation 
is given by the product of the dimensions of $\Lambda_{1}$ and $\Lambda_{2}$, 
times the dimension of $\mathbf{p}^{*}$.



Explicitly, the matrices corresponding to the zero-momentum block of the direct-product 
representations are
\begin{eqnarray}
M^{\left( \Lambda_{1}, \Lambda_{2}; \mathbf{p}^{*} \right)}_{ 
\alpha
\beta 
} \left( R \right)
=  
\left[ 
\Gamma^{\left( \Lambda_{1}; \mathbf{p}^{*} \right) }_{  \left( \alpha_{1}, \mathbf{p} \right), (\beta_{1}, \mathbf{p}^{'} )}
( R )
\right]
\left[
\Gamma^{\left( \Lambda_{2}; \mathbf{p}^{*} \right)}_{ \left( \alpha_{2}, -\mathbf{p} \right), (\beta_{2}, -\mathbf{p}^{'})}
( R )
\right],
\end{eqnarray}
where, on the left-hand side, we use $\alpha$, $\beta$ to denote 
the composite row and column indices $ \left(\alpha_{1}, \alpha_{2}, \mathbf{p} \right) $ 
and $( \beta_{1}, \beta_{2}, \mathbf{p}' )$.
In analogy to Eq.~\ref{eqn:irrep_project}, we can define a projection matrix
\begin{eqnarray}
  P^{\Lambda \lambda}_{\alpha \beta} = \frac{d_{\Lambda}} {n_{O_{h}^{D}}}
  \sum_{R \in O_{h}^{D}} \Gamma^{\left(\Lambda \right)}_{\lambda \lambda} \left( R \right)
M^{ \left( \Lambda_{1}, \Lambda_{2}; \mathbf{p}^{*} \right) }_{\beta \alpha} \left( R \right)^{*}.
\end{eqnarray}
Hence, if $\Lambda$ appears in the decomposition of $ \left[ \Lambda_{1}, \Lambda_{2}; \mathbf{p}^{*} \right] $,
and given the single-particle irreducible bases  $\left \{ O^{(\Lambda_{1})} \right \}$, 
$ \left \{ O^{(\Lambda_{2})} \right \} $,
the composite operator
\begin{eqnarray}
P^{\Lambda \lambda}_{\alpha \beta}  O^{\left(\Lambda_{1} \right)}_{\beta_{1}} ( t, \mathbf{p}{'} )
O^{\left(\Lambda_{2} \right)}_{\beta_{2}} ( t, -\mathbf{p}{'} )
\end{eqnarray}
(summation over $\beta$, i.e., $\beta_{1}$, $\beta_{2}$ and $\mathbf{p}'$, implied)
has zero net momentum, and transforms according to the row $\lambda$ of the 
octahedral-group irrep $\Lambda$.
The multiplicity of the irrep $\Lambda$ in the decomposition of $\left[ \Lambda_{1}, \Lambda_{2} ; \mathbf{p}^{*} \right]$
is given by the rank of the projection matrix $P^{\Lambda \lambda}$, which is simply the trace of $P^{\Lambda \lambda}$.
Averaging over all rows in $\Lambda$ gives the multiplicity in terms of the characters of $\Lambda$ and 
the little-group irreps:
\begin{eqnarray}
  n_{\Lambda} = \frac{1} {n_{G_{\mathbf{p}}^{D}}}
  \sum_{R \in G_{\mathbf{p}}^{D}} 
\left \{
\chi^{\left(\Lambda_{1}\right)} (R) 
\chi^{\left(\Lambda_{2}\right)} (R) 
\chi^{\left( \Lambda \right) *} \left( R \right)
\right \}.
\end{eqnarray}
Using this formula, one finds, for example, that the six-dimensional representation 
$ \left[ A_{2}, A_{2}; (0,0,1)^{*} \right]$, relevant 
for two pions with on-axis momenta, decomposes into 
$A_{1g} \oplus E_{g} \oplus T_{1 u}$,
and the twelve-dimensional pion-nucleon representation $[A_{2}, G_{1}; (0,0,1)^{*}]$ 
contains the $G_{1 g}$, $G_{1 u}$, $H_{g}$, $H_{u}$ irreps of 
$O_{h}$.

\subsection*{Work in progress}
To date, we have pruned single-hadron operators for all light-baryon, isovector-meson, isoscalar-meson, 
and kaon zero-momentum and non-zero momentum irreps.
Two-particle coefficients for all zero-momentum irreps have been computed, and measurements of the single-hadron constituent components have been performed on lattices with estimated $(3 {\rm{fm}})^{3}$ and $(4 {\rm{fm}})^{3}$ spatial volumes and a pion mass of approximately 240~MeV. Results for the stationary-state spectra on these lattices will be presented in the near future.

This work was supported by the U.S. NSF
under awards PHY-0510020, PHY-0653315,
PHY-0704171, PHY-0969863, and PHY-0970137, and
through TeraGrid/XSEDE resources provided by the
Pittsburgh Supercomputer Center, the Texas 
Advanced Computing Center, and the National Institute for Computational Sciences under grant numbers
TG-PHY100027 and TG-MCA075017.

\end{document}